\documentstyle[12pt,epsf]{article}
\catcode`\@=11
\def\maketitle{\par
 \begingroup
 \def\thefootnote{\fnsymbol{footnote}}
 \def\@makefnmark{\hbox
 to 0pt{$^{\@thefnmark}$\hss}}
 \if@twocolumn
 \twocolumn[\@maketitle]
 \else \newpage
 \global\@topnum\z@ \@maketitle \fi\thispagestyle{empty}\@thanks
 \endgroup 
 \setcounter{footnote}{0}
 \let\maketitle\relax
 \let\@maketitle\relax
 \gdef\@thanks{}\gdef\@author{}\gdef\@title{}\let\thanks\relax}
\def\@maketitle{\newpage
 \null
 \hbox to\textwidth{\hfil\hbox{\begin{tabular}{r}\@preprint\end{tabular}}}
 \vskip 2em \begin{center}
 {\large\bf \@title \par} \vskip 1.5em {\normalsize \lineskip .5em
\begin{tabular}[t]{c}\@author
 \end{tabular}\par}
 \end{center}
 \par
 \vskip 1.5em}
\def\preprint#1{\gdef\@preprint{#1}}
\def\abstract{\if@twocolumn
\section*{abstract}
\else \normalsize
\begin{center}
{\large\bf abstract\vspace{-.5em}\vspace{0pt}}
\end{center}
\quotation
\fi}
\def\endabstract{\if@twocolumn\else\endquotation\fi}
\catcode`\@=12

\begin{document}
\baselineskip=.285in
\preprint{SNUTP-96-81}


\title{\bf\large Chiral Corrections to the Axial Charges of \\
the Octet Baryons from Quenched QCD}
\author{\normalsize Myunggyu Kim and Seyong Kim\\[2mm]
{\normalsize\it Center for Theoretical Physics, Seoul National University,
Seoul 151-742, Korea}\\[2mm]
{\normalsize\it skim$@$ctp.snu.ac.kr}\\[2mm]
{\normalsize\it mgkim$@$galatica.snu.ac.kr}} 
\date{}
\maketitle

\begin{center}
{\large\bf Abstract} \\[3mm]
\end{center}
\indent\indent We calculate one-loop correction to the axial charges 
of the octet baryons using quenched chiral perturbation theory,
in order to understand chiral behavior of the axial charges in quenched
approximation to quantum chromodynamics (QCD). In contrast to regular
behavior of the full QCD chiral perturbation theory result,
$c_0+c_{l2}m_\pi^2\,\ln{m_\pi^2}+\cdots$, we find that the quenched chiral
perturbation theory result,
$c_0^Q+(c_{l0}^Q+c_{l2}^Qm_\pi^2)\ln{m_\pi^2}+c_2^Q 
m_\pi^2+\cdots$, is singular in the chiral limit.

\vspace{2mm}


\newpage

\pagenumbering{arabic}
\thispagestyle{plain}
Lattice quantum chromodynamics (QCD) simulation allows us to
investigate low energy phenomena of the strong interaction
using first principles of quantum field theory. It is well suited 
for understanding non-perturbative nature of the strong interaction. 
Thus far, this method has been successfully employed in calculating various 
low energy QCD related quantities \cite{Lat95}.

However, many of physical observables calculated in lattice QCD
use so-called ``quenched approximation'' in which vacuum polarization 
effects coming from quark--antiquark pair creation 
annihilation is neglected. This approximation is used due to the 
extensive computational cost in lattice simulation of full QCD, in
particular, of light quark system. Such truncation, otherwise in a first 
principle calculation, can cause undesirable effects on what we are 
interested in. It is often difficult to estimate such effects 
quantitatively. 
In addition to this, most cases of the current lattice
simulations are done with bare quark masses which produce pion masses
heavier than experimental value. Extrapolation of heavier mass results
from simulations to experimental pion mass region is necessary. Based
on the observation that one may use Taylor expansion for the mass 
dependence of physical quantities when the mass is small, a linear or 
quadratic fit has been used for the chiral extrapolation. One may 
question wisdom of such an extrapolation under the quenched approximation.

In order to understand quenching effects on chiral limit of
various physical observables in light meson system, recently, Sharpe 
devised a rule + digrammatic method \cite{Sharpe}. Bernard and 
Goltermann \cite{pBG} developed a systematic way based on the symmetry of 
quenched QCD (QQCD). The latter method is in similar spirit to chiral 
perturbation theory and an extension of the same idea to baryonic system 
is constructed by Labrenz and Sharpe \cite{pLS}. Using their
method, they found that the chiral behaviors of many physical 
quantities in the QQCD are
different from those in the full QCD and there are indeed sickness caused by
quenched approximation. The question is whether such
differences are numerically noticeable in physical pion mass region. 
If at the physical mass region, the deviation is small, we may safely
neglect it. Numerical investigations on meson mass (pion) revealed 
that although there is difference between the chiral behavior of the 
full QCD and that of 
the QQCD, the magnitude is small at the physical pion mass region 
\cite{Kim_Sinclair}. Similar conclusion may be drawn on the nucleon 
mass \cite{Sharpe2}. However, here, using the method developed in 
\cite{pLS}, we suggest that the chiral behavior of the axial charge of
the octet baryons in QQCD should depart noticeably from that of full QCD in the
physical pion mass region and in bare quark mass region currently used
in lattice QCD simulation. In this case, a quenched calculation may not
be trusted, let alone a linear extrapolation of it.

In the following, we briefly mention the most important ingredients
of the quenched chiral perturbation theory (Q$\chi$PT). Details are
available in \cite{pBG,pLS} and we follow the notations in \cite{pLS}.
Quenched approximation amounts to neglecting the quark determinant. 
Cancelling the quark determinant can be achieved by introducing additional
bosonic degrees of freedom, $\tilde{q}_i$, corresponding to each flavor of 
the quark field $q_i$. Each of $\tilde{q}_i$ has
the same mass, charge {\it etc} as the original quark. Since new
bosonic degrees of freedom has the same quantum number as fermion
degrees of freedom (quarks) except the spin, Gaussian integral
over the new degrees of freedom matches exactly the determinant
from the quark degrees of freedom with opposite power. Due to new
bosonic degrees of freedom, then, the symmetry of QQCD system becomes
the graded symmetry $U(3|3)\times U(3|3)$ for the quenched system
of light quarks, (u,d,s). This symmetry determines the form of the 
interactions among the pseudoscalar mesons in Q$\chi$PT. 

The dynamics of the mesons is conveniently described by an Hermitian
$6\times 6$ matrix field $\Phi$. A component $\Phi_{ij}$ has the same
transformation properties as the operator $Q_i\overline{Q}_j$ where
$Q=(u,d,s,\tilde{u},\tilde{d},\tilde{s})$. As in $\chi$PT (chiral
perturbation theory), the following is useful in constructing the Lagrangian:
\begin{equation}
\Sigma (x) = e^{2i\Phi(x)/f}, \xi(x) = e^{i\Phi(x)/f},
\end{equation}
\begin{equation}
A_\mu (x)=\frac{i}{2}(\xi\partial_\mu\xi^{\dag}-\xi^{\dag}\partial_\mu\xi),\,
V_\mu (x)=\frac{1}{2}(\xi\partial_\mu\xi^{\dag}+\xi^{\dag}\partial_\mu\xi).
\label{avcur}
\end{equation}
Under $U(3|3)\times U(3|3)$, the meson fields transform as
\begin{equation}
\Sigma\rightarrow L\Sigma R^{\dag},
\xi   \rightarrow L\xi U^{\dag}(x)=U(x)\xi R^{\dag}.
\end{equation}

The axial anomaly breaks this full chiral symmetry at the classical
level down to the semidirect product $[SU(3|3)\times SU(3|3)]\otimes U(1)$. 
The reduction in the symmetry introduces the field 
$\Phi_0=str(\Phi)/\sqrt{3}$ and allows the quenched chiral Lagrangian 
to include arbitrary functions of $\Phi_0$. The resulting Lagrangian 
in the mesonic sector is then
\begin{eqnarray}
{\cal L}^Q_{\phi}=\frac{f^2}{4}[str(\partial_\mu\Sigma\partial^\mu\Sigma^{\dag})
V_1(\Phi_0)+2\mu\,str(\xi^{\dag} m\xi + \xi m\xi^{\dag})V_2(\Phi_0)]
\nonumber\\
+\alpha_{\Phi}V_5(\Phi_0)\partial_\mu\Phi_0\partial^\mu\Phi_0
-m_0^2V_0(\Phi_0)\Phi_0^2,\label{mlag}
\end{eqnarray}
where $str$ denotes supertrace \cite{pBG,pLS} and 
$m=diag(m_u,m_d,m_s,m_u,m_d,m_s)$.
The potentials are normalized as $V_i(\Phi_0)=1+{\cal O}(\Phi_0^2)$.
In our calculations, the higher order terms in the potentials will not be
needed. In the full chiral theory, the vertex $i(\alpha_{\Phi}p^2-m_0^2)/3$
iterates infinitely in the $\eta'$ propagater. Thus the $\eta'$
acquires the heavy mass and can be
integrated out of the effective theory. In the quenched theory, however,
only one insertion of the vertex survives and thus the $\Phi_0$
remains light. We must therefore keep the last two terms in 
${\cal L}^q_{\Phi}$. Furthermore the lightness of the $\Phi_0$ will
introduce two additional parameters in the baryon sector as discused below.
The couplings, $\alpha_{\Phi}$ and $m_0^2$ can change usual power
counting rule in $\chi$PT if they are large. In loop calculation,
actual expansion parameters are $\alpha_{\Phi}/3$ and $m_0^2/3$ \cite{pBG}.
This leads one to expect that Q$\chi$PT power counting holds and we
calculate only leading behavior of these coupling.

We now define the fields for baryons. QQCD has $qq\tilde{q}$,
$q\tilde{q}\tilde{q}$, and $\tilde{q}\tilde{q}\tilde{q}$ baryons,
additional to the ususl $qqq$ baryons.
Spin-$1/2$ baryons belong to an irreducible representation of 
$SU(3|3)_v$ with dimension 70. $SU(3)$ decomposes it into  
an $8$ each of $qqq$'s and $\tilde{q}\tilde{q}\tilde{q}$'s and a
$1+8+8+10$ each of
$qq\tilde{q}$'s and $q\tilde{q}\tilde{q}$'s. The representation 
including  spin-$3/2$ baryons  
is 38 dimensional, which consisits of a $10$ of $qqq$'s, a $10$ and an
$8$ of $qq\tilde{q}$'s, an $8$ and a $1$ of $q\tilde{q}\tilde{q}$'s, 
and a $1$ of 
$\tilde{q}\tilde{q}\tilde{q}$'s. We call the baryons of the 70(38) simply
``octet''(``decuplet'') baryons. In Q$\chi$PT, the spin-1/2(3/2)
baryons are described by the tensor field
$B_{ijk}$($T^{\mu}_{ijk}$). They are defined to have
the same transformation properties as the following operators:
\begin{equation}
B^{\gamma}_{ijk}\sim [q^{\alpha,a}_{i}q^{\beta,b}_{j}q^{\gamma,c}_{k}
-q^{\alpha,a}_{i}q^{\gamma,c}_{j}q^{\beta,b}_{k}]
\varepsilon_{abc}(C\gamma_5)_{\alpha\beta},
\end{equation}
\begin{equation}
T^{\mu}_{\alpha,ijk}\sim [
q^{\alpha,a}_{i}q^{\beta, b}_{j}q^{\gamma,c}_{k}+
q^{\beta ,b}_{i}q^{\gamma,c}_{j}q^{\alpha,a}_{k}+
q^{\gamma,c}_{i}q^{\alpha,a}_{j}q^{\beta ,b}_{k}]
\varepsilon_{abc}(C\gamma_\mu)_{\beta\gamma}
\end{equation}
where $C=i\gamma_2\gamma_0$ is the charge conjugation matrix and
$a$, $b$ and $c$ are color indices. Both fields have the same transformation
properties, exemplified by
\begin{equation}
B_{\gamma,ijk}\rightarrow(-)^{i'(j+j')+(i'+j')(k+k')}
U_{ii'}U_{jj'}U_{kk'}B_{\gamma,i'j'k'},
\end{equation}
where $U\in SU(3|3)$. The grading factor stems from the fact that
the off-diagonal $3\times 3$ blocks of $U$ are grassman variables.
The notations for the indices are as follows: 1 for the anticommuting
variables ($i=1,2,3$) and 0 for the commuting variables($i=4,5,6$).
The baryon fields satisfy the symmetry properties
\begin{eqnarray}
B_{ijk}=(-)^{jk+1}B_{ikj},
\nonumber\\
B_{ijk}+(-)^{ij+1}B_{jik}+(-)^{ij+jk+ki+1}B_{kji},
\nonumber\\
T_{ijk}=(-)^{ij+1}T_{jik}=(-)^{jk+1}T_{ikj}.\label{bspro}
\end{eqnarray}  
 
The Lagrangian in the baryon sector at lowest order can be written 
in terms of the invariant bilinears of the baryon fields. 
They include the covariant derivative or the field $A^\mu$.
The covariant derivative of the field $B_{ijk}$ is
\begin{equation}
D^\mu B_{ijk}=\partial^\mu B_{ijk}+(V^\mu)_{ii'}B_{i'jk}
+(-)^{i(j+j')}(V^\mu)_{jj'}B_{ij'k}
+(-)^{(i+j)(k+k')}(V^\mu)_{kk'}B_{ijk'}
\end{equation}
where $V^\mu$ is a vector current defined by eq.~(\ref{avcur}). The
covariant derivarive of the field $T^\mu_{ijk}$ takes the same form.
We follow \cite{pLS}'s notations for the contraction of the flavor
indices. They are
\begin{equation}
\overline{C}\Lambda C'\equiv\overline{C}_{kji}\Lambda C_{ijk}',
\end{equation}
\begin{equation}
\overline{C}\Lambda E C'\equiv\overline{C}_{kji}\Lambda E_{ii'}C_{i'jk}',
\end{equation}
 \begin{equation}
\overline{C}\Lambda C' E\equiv\overline{C}_{kji}\Lambda E_{kk'}C_{ijk'}'
(-)^{(i+j)(k+k')},
\end{equation}
where $C$ and $C'$ are the baryon fields, $E$ is a matrix field,
and $\Lambda$ is an arbitrary Dirac matrix. An example of $E$ is $A^\mu$.

The lowest order Lagrangian is, then, 
\begin{equation}
{\cal L}^Q={\cal L}^Q_{\Phi}+{\cal L}^Q_{BT\Phi}
\end{equation}
where ${\cal L}^Q_{\Phi}$ is given in eq.~(\ref{mlag}) and
\begin{eqnarray}
{\cal L}^Q_{BT\Phi}=i\overline{B}v\cdot D B
+i\overline{T}^\nu v\cdot DT_\nu+\Delta M\overline{T}^\nu T_\nu \nonumber\\
+2\alpha\overline{B}S_\mu B A^\mu
+2\beta \overline{B}S_\mu A^\mu B
+2\gamma\overline{B}S_\mu B str(A^\mu) \nonumber\\
+2H  \overline{T}^{\nu} S_\mu A^\mu T_{\nu}
-\sqrt{\frac{3}{2}}C[  \overline{T}^{\nu} A^{\nu} B
+               \      \overline{B}       A_{\nu} T^{\nu}]
+2\gamma'\overline{T}^{\nu} S_\mu T_{\nu} str(A^\mu).
\end{eqnarray}
The symbols $v$ and $S^\mu$ denote the fixed velocity and the spin
operator in the heavy baryon formalism~\cite{pGeorgi}. 
Various other possible terms
such as $\overline{B}_{ijk}S^{\mu}A_{\mu,jj'}B_{ij'k}(-)^{i(j+j')}$,
{\it etc.} are not independent due to the symmetry properties
given in eq.~(\ref{bspro}). In caculating the contributions 
of decuplet intermediate baryons,
we assume that $\Delta m \ll m_\pi$ and treat the octet and the
decuplet baryons as degenerate. 

When the flavor indices are restricted to the range $1-3$, the baryon
fields in the quenched theory are explicitely related to ones in 
the full theory as
\begin{equation}
B_{ijk}\mid_R=\frac{1}{\sqrt{6}}(\varepsilon_{ijk'}B_{k'k}^{full}
+\varepsilon_{ikk'}B_{k'j}^{full}),
\end{equation}
\begin{equation}
B_{ijk}\mid_R=T_{ijk}^{full}.
\end{equation}
Using these relations, one can formally show that the quenched Lagrangian
is equal to the full Lagrangian under the restrictions of the flavor
indices and the identifications of the parameters
\begin{equation}
  \alpha=2(\frac{1}{3}D+F),\, \beta=(-\frac{5}{3}D+F),
\,\gamma=2(D-F),\label{pide}
\end{equation}
although there is no reason why the parameters of two theories
should be related to each other. For better comparison with $\chi$PT
result, we will re-express our results using these relations except $\gamma$.
  
The axial current from the quenched chiral Lagrangian is given by
\begin{eqnarray}
J^a_\mu=i\frac{f^2}{2}str[T^a(\partial_\mu\Sigma^{\dag}\Sigma-
\partial_\mu\Sigma\Sigma^{\dag})]
+v_\mu[\overline{B}\Omega^A_{-}B+2\overline{B}B\Omega^A_{-} 
+3\overline{T}^{\nu}\Omega^A_{-} T_{\nu}]\nonumber\\
+2\alpha\overline{B}S_\mu B \Omega^A_{+}
+2\beta \overline{B}S_\mu\Omega^A_{+} B
+2\gamma\overline{B}S_\mu B str(\Omega^A_{+})\nonumber\\
+2H       \overline{T}^{\nu} S_\mu\Omega^A_{+} T_{\nu}
+2\gamma' \overline{T}^{\nu} S_\mu T_{\nu}str(\Omega^A_{+}) 
-\sqrt{\frac{3}{2}}C(\overline{T}^{\mu}\Omega^A_{+}B
                    +\overline{B}      \Omega^A_{+}T^{\mu}),\label{axial}
\end{eqnarray}
\begin{equation}
\Omega^A_{\pm}=\frac{1}{2}(\xi T^A\xi^{\dag}\pm \xi^{\dag} T^A\xi),\,
\end{equation}
\begin{equation}
T^A=\frac{1}{2}
\left[\begin{array}{cc}
\lambda^A & 0 \\ 0 & \lambda^A
\end{array}\right],
\end{equation}
where $\lambda^A$ are Gell-Mann matrices. The numerical factors 2 and 3
in the second line is explained by the symmetry properties given 
in eq.~(\ref{bspro}).

\begin{figure}
\vspace{9pt}
\epsfxsize=70mm
\leavevmode
\hspace{3cm}\epsffile[50 50 410 302]{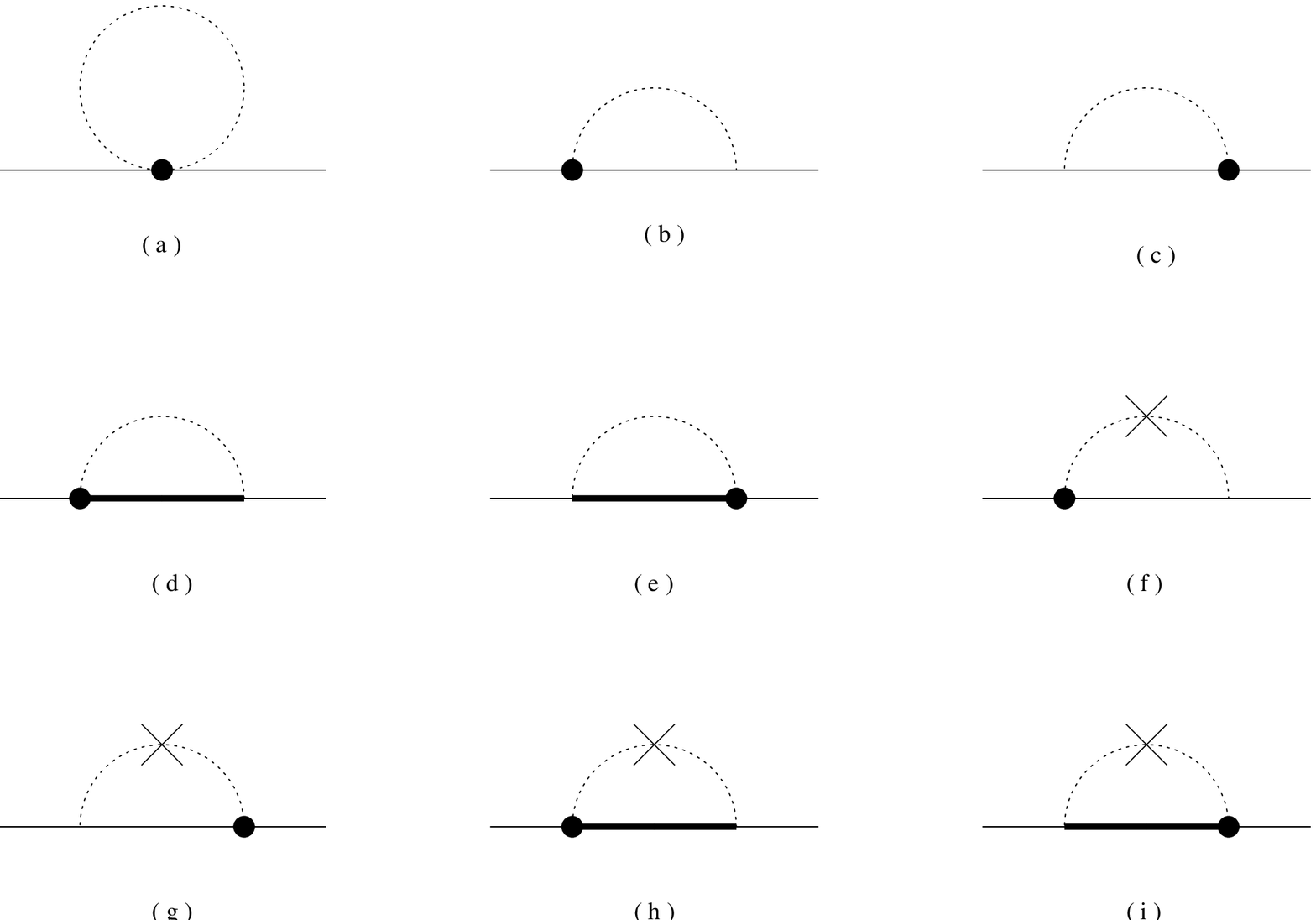}\vspace{1cm}
\caption{Vanishing graphs. The solid dots represent the vertices
  steming from the axial current.}
\label{figv}
\end{figure}

\begin{figure}
\vspace{9pt}
\epsfxsize=80mm
\leavevmode
\hspace{3cm}\epsffile[50 50 410 302]{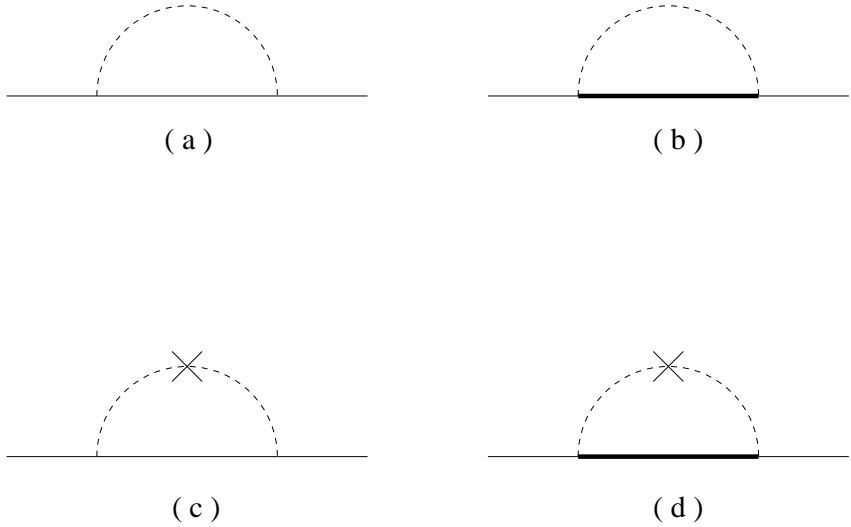}\vspace{-1cm}
\caption{Wavefunction renormalization graphs.
The hairpin represents the vertex $i(\alpha_{\Phi}p^2-m_0^2)/3$.}
\label{figw}
\end{figure}

\begin{figure}
\vspace{9pt}
\epsfxsize=70mm
\leavevmode
\hspace{3cm}\epsffile[50 50 410 302]{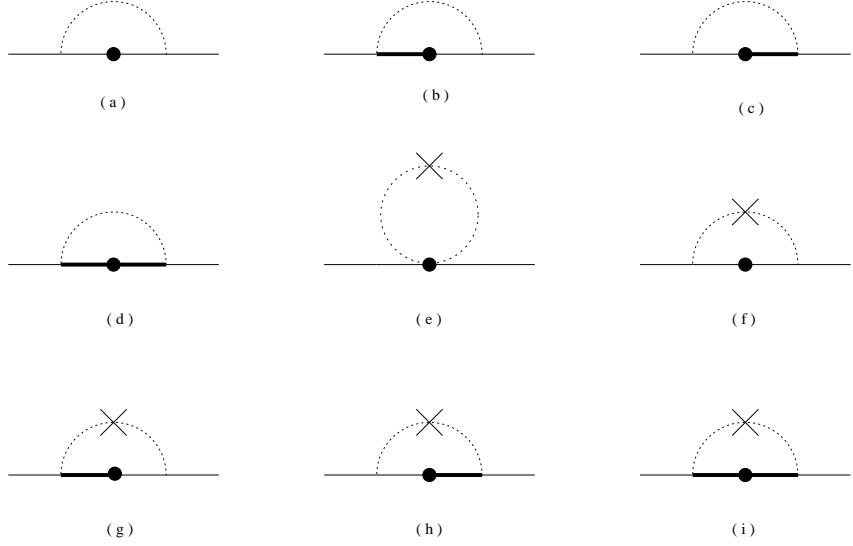}\vspace{1cm}
\caption{Graphs contributing to the coefficients 
$\alpha^{A}_{ij,ab}$ and $\beta^{A}_{ij,ab}$.
The solid dots represent the vertices steming from the axial
current. The hairpin represents the vertex $i(\alpha_{\Phi}p^2-m_0^2)/3$.}
\label{figab}
\end{figure}

The renormalization of the axial currents can be done by
computing the diagrams given in the figures~\ref{figv}, \ref{figw} 
and \ref{figab} using Feynman rules derived from ${\cal L}_\Phi^Q$ and
${\cal L}_{BT\Phi}^Q$ and the vertices from eq.~\ref{axial}. 
All the diagrams in the figure~\ref{figv} vanish:
(a) vanishes since only closed quark loops are present (which cannot be
present in the quenched approximation) and the others
are zero due to the property $v\cdot S=0$.  
Then, the matrix element of $J^A$ for the octet baryons $B_i$ and $B_j$
can be written in the form 
\begin{eqnarray}
\langle B_i\mid J^A \mid B_j\rangle=
\overline{u}_{B_i}\gamma^{\mu}\gamma^{5} u_{B_j}\,t^A_{ij}[1
+\sum_{a\leq b}(\alpha^A_{ij,ab}-\lambda_{ij,ab})X_{ab}
+\sum_{a\leq b}(\beta ^A_{ij,ab}-\rho_{ij,ab})Y_{ab}],\label{resulteq}
\end{eqnarray}
\begin{equation}
X_{ab}=\frac{M_{ab}^2}{16\pi^2 f^2}\ln{\frac{M_{ab}^2}{\mu^2}},
\end{equation}
\begin{equation}
Y_{ab}=\frac{(\alpha_{\Phi} M_{aa}^2-m_0^2)X_{aa}
-(\alpha_{\Phi} M_{bb}^2-m_0^2)X_{bb}}
{3(M_{aa}^2-M_{bb}^2)},
\end{equation}
\begin{equation}
Y_{aa}=\frac{1}{48\pi^2 f^2}[(2\alpha_{\Phi} M_{aa}^2-m_0^2)
\ln{\frac{M_{aa}^2}{\mu^2}}+(\alpha_{\Phi} M_{aa}^2-m_0^2)],
\end{equation}
where $t^A_{ij}$ is the tree level result, $\lambda_{ij,ab}$ and
$\rho_{ij,ab}$ are the wavefunction renormalization factors
without and with the hairpin vertex, and 
$\alpha^A_{ij,ab}$ and $\beta^A_{ij,ab}$ are the one-loop corrections 
without and with the hairpin vertex.
We list the coefficients at the end of this article.  
Here we use $m_u=m_d$.

The $\mu$ is the scale introduced in the dimensional
regulazation. The scale dependence is cancelled by the counter terms
obtained by the $O(m_q)$ Lagragians such as
\begin{equation}
\frac{(\mu^2)^{4-d}}{\Lambda_\chi}\bar{B}(\xi^\dagger m\xi^\dagger+\xi m\xi)
S^{\mu}A_{\mu}B
\end{equation}
where $m$ is the quark mass matrix. Including the counter terms, the function
$X_{ab}$ changes to
\begin{equation}
X_{ab}\rightarrow
\frac{M_{ab}^2}{16\pi f^2}[\ln\frac{M_{ab}^2}{\mu^2}+c(\mu)]
\end{equation}
where $c(\mu)$ is the finite part of the counter term. We take
$\mu\simeq\Lambda_{\chi}\sim\,1\,GeV$ 
and neglect the finite part. This is fully
justified  near the chiral limit since $|\ln(M_{ab}^2/\mu^2)|\gg
c(\mu)$. In the real world the logarithmic correction is not
significantly large but we expect it to give an estimate of the actual size
of the corrections~\cite{Luty}.

It is interesting to compare the full and the quenched chiral
corrections to one of the above matrix elements. We choose
the isovetor axial charge of the proton, $g_A$ for this.
It is equal to the $\langle p|J^{1+i2}|n\rangle$ due to the isospin symmetry.
Quenched lattice simulation data for this quantity is also 
availiable \cite{KFLiu}.
Jenkins and Manohar computed full chiral logarithmic contributions to the
baryon axial vector currents including both intermediate octet and
decuplet states~\cite{pJM,pJM2}. Their full chiral result for $g_A$ is
\begin{eqnarray}
g_A=(D+F)[1-\{1+2(D+F)^2+\frac{2C^2(9D+9F+25H)}{81(D+F)}\}
\frac{M_{\pi}^2}{16\pi^2 f^2}\ln{\frac{M_{\pi}^2}{\mu^2}}
\nonumber\\
-\{\frac{1}{2}+\frac{15D^3-D^2F+3DF^2+33F^3}{6(D+F)}
+\frac{C^2(-27D+45F+20H)}{162(D+F)}\}
\frac{M_{K}^2}{16\pi^2 f^2}\ln{\frac{M_{K}^2}{\mu^2}}
\nonumber\\
-\frac{(-D+3F)^2}{3}
\frac{M_{\eta}^2}{16\pi^2 f^2}\ln{\frac{M_{\eta}^2}{\mu^2}}].
\end{eqnarray}
On the other hand, our result for the quenched $g_A$ is
\begin{eqnarray}
g_A^Q=(D+F)[1+\{\frac{4(D-3F)(D^2+2DF+3D\gamma+3F\gamma)}{3(D+F)}
\nonumber \\
+\frac{C^2(-15D+9F-10H)}{27(D + F)}\}\frac{M_{\pi}^2}{16\pi^2 f^2}
\ln{\frac{M_{\pi}^2}{\mu^2}} \nonumber \\
+2(-D + 3F)^2 \frac{1}{48\pi^2 f^2}\{(2\alpha_{\Phi} M_{\pi}^2-m_0^2)
\ln{\frac{M_{\pi}^2}{\mu^2}}+(\alpha_{\Phi} M_{\pi}^2-m_0^2)\}].
\end{eqnarray}

In full theory, having only octet states, there are two free parameters
$D$ and $F$ in their expressions. A $\chi^2$ fit to the hyperon
semileptonic decays gives $D\sim 0.56$ and $F\sim 0.33$. Inclusion of
the intermediate decuplet states needs two additional parameters
$C^2$ and $H$. Jenkins and Manohar determined the parameter $C^2$ by 
fitting the $\Delta\rightarrow N\pi$ decay rate, $\mid C\mid\sim
1.6$. The three 
parameter fit yielded $D\sim 0.61$, $F\sim 0.40$ and $H\sim -1.9$.
They used $\mu\sim 1\, GeV$.

The quenched chiral expressions for the axial currents given by
eq.(\ref{resulteq}) need seven parameters $D$, $F$, $C^2$, $H$, $\gamma$,
$\alpha_{\Phi}$ and $m_0$. One way to determine the values of the parameters
is to fit to the quenched lattice data for the axial currents obtained
on the physical quark masses. However, such simulation data is not yet
available at present and we are interested in
the qualitative comparison between the full chiral and the quenched chiral 
behaviors of the physical quantities rather than actual values of the 
parameters. Thus, for the sake of comparison, we use the values for the
parameters 
$D$, $F$, $C^2$ and $H$ obtained by Jenkins and Manohar. 
We use the physical $\eta'$ mass for the value of 
the parameter $m_0$. 
We determine
the value of the parameter $\gamma$ by the relation $\gamma=2(D-F)$
and choose $\alpha_{\Phi}=0$. Our study shows that the variation of 
our result is less than 10 \% for the values $\gamma=0\sim 1$ and
the dependence on $\alpha_\Phi$ is much weaker (the variation
less than 3 \% for $\alpha_\Phi=-1\sim+1$). 

\begin{figure}
\vspace{9pt}
\epsfxsize=70mm
\leavevmode
\hspace{2.2cm}\epsffile{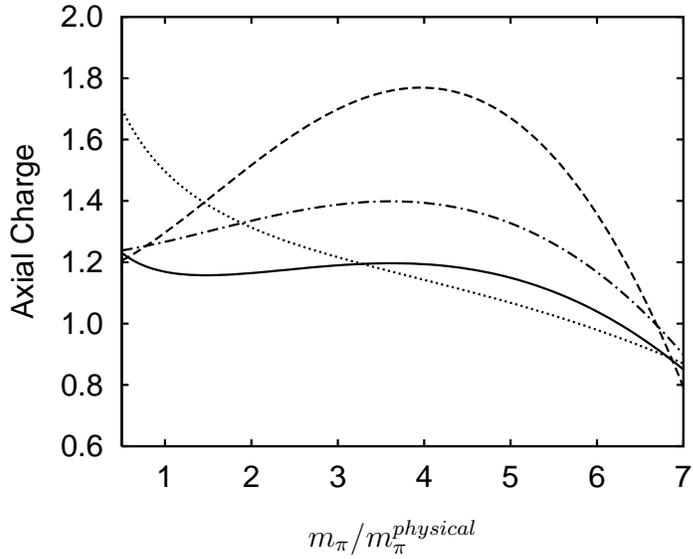}\vspace{-0.2cm}
\centerline{$m_{\pi}/m_{\pi}^{physical}$}
\caption{Chiral behaviors of the nucleon isovector axial charge
from Q$\chi$PT and $\chi$PT
with only octet intermediate states and both octet and decuplet states. 
The solid, dotted, dashed, and dot-dashed lines
show the chiral behaviors 
form Q$\chi$PT with octets, Q$\chi$PT with both octets and decuplets, 
 $\chi$PT with octets, and $\chi$PT with both octets and decuplets, 
in order.}
\label{figplot}
\end{figure}

In the figure~\ref{figplot}, we show the full and the quenched chiral behaviors
of the nucleon isovector axial charge with only octet states and 
both octet and decuplet states using the parameter values mentioned in
the above (in the full chiral calculations, 
the $K$-$\pi$ and the $\eta$-$K$ mass differences are fixed at 
the physical values). In the quenched chiral 
calculations the meson loops with the $s$ flavor do not contribute.
Full $\chi$PT result depends on pion mass
visibly. However, in the region $4\, m_\pi^{physical}$ 
and $6\, m_\pi^{physical}$, the two plots of the $g_A^Q$ are almost 
flat with flatter behavior for octet states only. 
Of course these results are reliable only 
for sufficieltly small pion mass because the higher order corrections 
become important as the pion mass increases and we should take this figure 
as only an indication. There exists a quenched lattice simulation 
data of the 
axial charge \cite{KFLiu} but due to the reason stated in the above, 
we can not compare our one loop result with this lattice data.

In short, one can see that in the limit $m_\pi\rightarrow 0$,
\begin{equation}
g_A \rightarrow (D+F),\,g_A^Q\rightarrow c^Q \ln{m_\pi}.
\end{equation} 
This singular behavior of the axial charge of the octet baryons,
$g_A^Q$, is due to the quenched approximation to QCD. Similar to the case
cosidered in observables of heavy-light meson system \cite{Sharpe_Zhang},
the quenched artifact in the axial charge occurs as a
leading singularity.

Finally we give the expressions for the coefficients appearing 
in eq.~\ref{resulteq}. 
The expressions for the $t^A_{ij}$ are
\begin{eqnarray}
t^{8}_{pp}=
{{-D + 3\,F}\over {2\,{\sqrt{3}}}},
\,
t^{1+i2}_{pn}=
D + F,
\,
t^{1+i2}_{\Lambda\Sigma^{-}}=
{\sqrt{{2\over 3}}}\,D,
t^{1+i2}_{\Xi^{0}\Xi^{-}}=
D - F,
\nonumber\\
t^{4+i5}_{p\Lambda}=
-{{D + 3\,F}\over {{\sqrt{6}}}},
\,
t^{4+i5}_{\Lambda\Xi^{-}}=
{{-D + 3\,F}\over {{\sqrt{6}}}},
\,
t^{4+i5}_{n\Sigma^{-}}=
D - F,
\,
t^{4+i5}_{\Sigma^{0}\Xi^{-}}=
{{D + F}\over {{\sqrt{2}}}}
=\frac{1}{\sqrt{2}}t^{4+i5}_{\Sigma^{+}\Xi^{0}}.
\end{eqnarray}

The coefficients $\lambda_{ij,ab}$ and
$\rho_{ij,ab}$ are defined by
\begin{eqnarray}
&&\lambda_{ij,ab}=(\lambda_{i,ab}+\lambda_{j,ab})/2,\nonumber\\
&&\rho_{ij,ab}=(\rho_{i,ab}+\rho_{j,ab})/2.
\end{eqnarray}
The diagrams (a) and (b) in the figure~\ref{figw} yield
\begin{eqnarray}
&&\lambda_{\Lambda,uu}=
-\frac{1}{3}D^2+4DF-3F^2-(4D-6F)\gamma+\frac{1}{2}C^2
,\nonumber\\
&&\lambda_{\Lambda,us}=-5D^2+2DF+3F^2+\frac{1}{2}C^2,\,\,\,
\lambda_{\Lambda,ss}=(D+3F)\gamma,
\end{eqnarray}
\begin{eqnarray}
&&\lambda_{\Sigma,uu}=-D^2+3F^2+6F\gamma+\frac{1}{6}C^2
,\nonumber\\
&&\lambda_{\Sigma,us}=-D^2+6DF-3F^2+\frac{5}{6}C^2,\,\,\,
\lambda_{\Sigma,ss}=3(-D+F)\gamma,
\end{eqnarray}
\begin{eqnarray}
&&\lambda_{N,uu}=\lambda_{\Sigma,uu}+\lambda_{\Sigma,us}+\lambda_{\Sigma,ss}
,\,\,\,\lambda_{N,us}=\lambda_{N,ss}=0,
\end{eqnarray}
\begin{eqnarray}
\lambda_{\Xi,uu}=\lambda_{\Sigma,ss},\,\,\,
\lambda_{\Xi,us}=\lambda_{\Sigma,us},\,\,\,
\lambda_{\Xi,ss}=\lambda_{\Sigma,uu}.
\end{eqnarray}
The diagrams (c) and (d) in the figure~\ref{figw} yield
\begin{eqnarray}
\rho_{\Lambda,uu}=
-\frac{3}{2}(\frac{4}{3}D-2F)^2,\,\,\,
\rho_{\Lambda,us}=3(\frac{4}{3}D-2F)(\frac{1}{3}D+F),\,\,\,
\rho_{\Lambda,ss}=-\frac{3}{2}(\frac{1}{3}D+F)^2\gamma,
\end{eqnarray}
\begin{eqnarray}
\rho_{\Sigma,uu}=-6F^2-\frac{1}{3}C^2,\,\,\,
\rho_{\Sigma,us}=6F(D-F)+\frac{2}{3}C^2,\,\,\,
\rho_{\Sigma,ss}=-\frac{3}{2}(D-F)^2-\frac{1}{3}C^2,
\end{eqnarray}
\begin{eqnarray}
\rho_{N,uu}=
\rho_{\Sigma,uu}+\rho_{\Sigma,us}+\rho_{\Sigma,ss}
,\,\,\,\rho_{N,us}=\rho_{N,ss}=0,
\end{eqnarray}
\begin{eqnarray}
\rho_{\Xi,uu}=\rho_{\Sigma,ss},\,\,\,
\rho_{\Xi,us}=\rho_{\Sigma,us},\,\,\,
\rho_{\Xi,ss}=\rho_{\Sigma,uu}.
\end{eqnarray}

 The $\alpha^A_{ij,ab}$ corresponds to the digrams without
the hairpin in the figure~\ref{figab}.
Our result for the $\alpha^A_{ij,ab}$ is
\begin{eqnarray}
&&\alpha^{1+i2}_{pn,uu}=
{{2\,D\,\left( D - 3\,F \right) \,
      \left( -D + F \right) }\over {3\,\left( D + F \right) }} + 
  \left( D - 3\,F \right) \,\gamma 
+
{{2\,{{C}^2}\,\left( 6\,D + 18\,F - 5\,H \right) }\over 
   {27\,\left( D + F \right) }}
,\nonumber\\ 
&&\alpha^{1+i2}_{pn,us}=\alpha^{1+i2}_{pn,ss}=0
,
\end{eqnarray}
\begin{eqnarray}
&&\alpha^{1+i2}_{\Lambda\Sigma^{-},uu}=
{{4\,D\,\left( -D + 3\,F \right) }\over 9} + 
  {{2\,\left( D - 3\,F \right) \,\gamma }\over 3}
+
{{{{C}^2}\,\left( -6\,D + 18\,F - 18\,\gamma  - 5\,H \right)}\over
     {54\,D}}
,\nonumber\\ 
&&\alpha^{1+i2}_{\Lambda\Sigma^{-},us}=
{{D\,\left( D - 3\,F \right) }\over 9}
+
{{{{C}^2}\,\left( 48\,D + 72\,F - 5\,H \right) }\over 
   {108\,D}}
,\nonumber\\ 
&&\alpha^{1+i2}_{\Lambda\Sigma^{-},ss}=
{{\left( D - 3\,F \right) \,\gamma }\over 3}
+
{{{{C}^2}\,\gamma }\over {3\,D}}
, 
\end{eqnarray}
\begin{eqnarray}
&&\alpha^{1+i2}_{\Xi^{0}\Xi^{-},uu}=
\left( D - F \right) \,\gamma 
+
{{-4\,{{C}^2}\,\gamma }\over {9\,\left( D - F \right) }}
,\nonumber\\ 
&&\alpha^{1+i2}_{\Xi^{0}\Xi^{-},us}=
{{-{{D}^3} + 3\,{{D}^2}\,F - 27\,D\,{F^2} + 9\,{F^3}}\over 
   {9\,\left( -D + F \right) }}
+
{{{{C}^2}\,\left( 18\,D + 18\,F + 5\,H \right) }\over 
   {81\,\left( D - F \right) }}
,\nonumber\\ 
&&\alpha^{1+i2}_{\Xi^{0}\Xi^{-},ss}=
{{{{D}^3} + 3\,{{D}^2}\,F + 9\,D\,{F^2} - 9\,{F^3}}\over 
    {9\,\left( -D + F \right) }} - 2\,F\,\gamma 
+
{{{{C}^2}\,\left( 72\,F + 72\,\gamma  + 5\,H \right) }\over 
   {162\,\left( D - F \right) }}
, 
\end{eqnarray}
\begin{eqnarray}
&&\alpha^{4+i5}_{p\Lambda,uu}=
{{{{D}^3} + 45\,{{D}^2}\,F - 81\,D\,{F^2} + 27\,{F^3}}\over 
    {18\,\left( D + 3\,F \right) }} + 
  {{\left( 7\,D - 15\,F \right) \,\gamma }\over 6}
+
{{{{C}^2}\,\left( 9\,D + 9\,F - 5\,H \right) }\over 
   {9\,\left( D + 3\,F \right) }}
,\nonumber\\ 
&&\alpha^{4+i5}_{p\Lambda,us}=
{{-25\,{{D}^3} + 63\,{{D}^2}\,F - 27\,D\,{F^2} - 
     27\,{F^3}}\over {18\,\left( D + 3\,F \right) }}
+
{{-\left( {{C}^2}\,\left( 6\,D - 18\,F + 5\,H \right)  \right)}
    \over {18\,\left( D + 3\,F \right) }}
,\nonumber\\ 
&&\alpha^{4+i5}_{p\Lambda,ss}=
{{-\left( \left( D + 3\,F \right) \,\gamma  \right) }\over 6}
,
\end{eqnarray}
\begin{eqnarray}
&&\alpha^{4+i5}_{\Lambda\Xi^{-},uu}=
{{\left( -D + F \right) \,
      \left( 5\,{{D}^2} - 6\,D\,F + 9\,{F^2} \right) }\over 
    {6\,\left( -D + 3\,F \right) }} + 
  {{\left( 7\,D - 9\,F \right) \,\gamma }\over 6}
+
{{{{C}^2}\,\left( -9\,D + 9\,F - 18\,\gamma  - 5\,H \right) }\over
     {27\,\left( -D + 3\,F \right) }}
,\nonumber\\ 
&&\alpha^{4+i5}_{\Lambda\Xi^{-},us}=
{{4\,D\,\left( {{D}^2} + 6\,D\,F - 9\,{F^2} \right) }\over 
   {9\,\left( -D + 3\,F \right) }}
+
{{2\,{{C}^2}\,\left( 3\,D - 9\,F - 5\,H \right) }\over 
   {27\,\left( -D + 3\,F \right) }}
,\nonumber\\ 
&&\alpha^{4+i5}_{\Lambda\Xi^{-},ss}=
{{\left( D + 3\,F \right) \,
      \left( -5\,{{D}^2} + 6\,D\,F - 9\,{F^2} \right) }\over 
    {18\,\left( -D + 3\,F \right) }} - 
  {{\left( D + 9\,F \right) \,\gamma }\over 6}
+
{{{{C}^2}\,\left( D + 3\,F + 6\,\gamma  \right) }\over 
   {9\,\left( -D + 3\,F \right) }}
, 
\end{eqnarray}
\begin{eqnarray}
&&\alpha^{4+i5}_{n\Sigma^{-},uu}=
{{{{D}^3} + 9\,{{D}^2}\,F - 9\,D\,{F^2} - 9\,{F^3}}\over 
    {18\,\left( -D + F \right) }} + 
  {{\left( D - 5\,F \right) \,\gamma }\over 2}
+
{{{{C}^2}\,\left( 9\,D + 45\,F + 18\,\gamma  + 5\,H \right) }\over
     {81\,\left( D - F \right) }}
,\nonumber\\ 
&&\alpha^{4+i5}_{n\Sigma^{-},us}=
{{-{{D}^3} + 3\,{{D}^2}\,F - 27\,D\,{F^2} + 9\,{F^3}}\over 
   {18\,\left( -D + F \right) }}
+
{{{{C}^2}\,\left( 18\,D + 18\,F + 5\,H \right) }\over 
   {162\,\left( D - F \right) }}
,\nonumber\\ 
&&\alpha^{4+i5}_{n\Sigma^{-},ss}=
{{\left( D - F \right) \,\gamma }\over 2}
+
{{-2\,{{C}^2}\,\gamma }\over {9\,\left( D - F \right) }}
, 
\end{eqnarray}
\begin{eqnarray}
&&\alpha^{4+i5}_{\Sigma^{0}\Xi^{-},uu}=
{{\left( D - F \right) \,\left( -{{D}^2} + 6\,D\,F + 
        3\,{F^2} \right) }\over {6\,\left( D + F \right) }} + 
  {{\left( D - 3\,F \right) \,\gamma }\over 2}
+
{{{{C}^2}\,\left( -9\,D + 9\,F - 5\,H \right) }\over 
   {81\,\left( D + F \right) }}
,\nonumber\\ 
&&\alpha^{4+i5}_{\Sigma^{0}\Xi^{-},us}=
{{\left( -D + F \right) \,\left( {{D}^2} + 3\,{F^2} \right) }\over
     {3\,\left( D + F \right) }}
+
{{2\,{{C}^2}\,\left( 27\,D + 45\,F - 10\,H \right) }\over 
   {81\,\left( D + F \right) }}
,\nonumber\\ 
&&\alpha^{4+i5}_{\Sigma^{0}\Xi^{-},ss}=\alpha^{4+i5}_{\Sigma^{0}\Xi^{-},uu}
, 
\end{eqnarray}
\begin{eqnarray}
&&\alpha^{4+i5}_{\Sigma^{+}\Xi^{0},ab}=\alpha^{4+i5}_{\Sigma^{0}\Xi^{-},ab}
,
\end{eqnarray}
\begin{eqnarray}
&&\alpha^{8}_{pp,uu}=
{{2\,D\,\left( D - 3\,F \right) }\over 3} + 
  \left( D - 3\,F \right) \,\gamma 
+
{{5\,{{C}^2}\,H}\over {9\,\left( D - 3\,F \right) }}
,\,\,\,
\alpha^{8}_{pp,us}=\alpha^{8}_{pp,ss}=0
. 
\end{eqnarray}

The diagrams with the hairpin in the figure~\ref{figab} yield the
results for $\beta^A_{ij,ab}$ which are
\begin{eqnarray}
&&\beta^{1+i2}_{pn,uu}=
{{{{\left( -D + 3\,F \right) }^2}}\over 2}
,\,\,\, 
\beta^{1+i2}_{pn,us}=\beta^{1+i2}_{pn,ss}=
0
,
\end{eqnarray}
\begin{eqnarray}
&&\beta^{1+i2}_{\Lambda\Sigma^{-},uu}=
{{2\,F\,\left( -2\,D + 3\,F \right) }\over 3}
+
{{2\,{{C}^2}\,\left( -2\,D + 3\,F \right) }\over {9\,D}}
,\nonumber\\ 
&&\beta^{1+i2}_{\Lambda\Sigma^{-},us}=
{{2\,\left( {{D}^2} - 2\,D\,F + 3\,{F^2} \right) }\over 3}
+
{{{{C}^2}\,\left( 5\,D - 3\,F \right) }\over {9\,D}}
,\nonumber\\ 
&&\beta^{1+i2}_{\Lambda\Sigma^{-},ss}=
{{-{{D}^2}}\over 6} - {{D\,F}\over 3} + {{{F^2}}\over 2}
+
{{-\left( {{C}^2}\,\left( D + 3\,F \right)  \right) }\over 
   {9\,D}}
,
\end{eqnarray}
\begin{eqnarray}
&&\beta^{1+i2}_{\Xi^{0}\Xi^{-},uu}=
{{{{\left( D - F \right) }^2}}\over 2}
+
{{{{C}^2}\,\left( -36\,D + 36\,F - 5\,H \right) }\over 
   {81\,\left( D - F \right) }}
,\nonumber\\ 
&&\beta^{1+i2}_{\Xi^{0}\Xi^{-},us}=
-2\,\left( D - F \right) \,F
+
{{2\,{{C}^2}\,\left( 18\,D + 18\,F + 5\,H \right) }\over 
   {81\,\left( D - F \right) }}
,\nonumber\\ 
&&\beta^{1+i2}_{\Xi^{0}\Xi^{-},ss}=
2\,{F^2}
+
{{-\left( {{C}^2}\,\left( 72\,F + 5\,H \right)  \right) }\over 
   {81\,\left( D - F \right) }}
,
\end{eqnarray}
\begin{eqnarray}
&&\beta^{4+i5}_{p\Lambda,uu}=
{1\over 4}
+
{{2\,{{D}^2}}\over 3} - 3\,D\,F + 3\,{F^2}
,\nonumber\\ 
&&\beta^{4+i5}_{p\Lambda,us}=
-{1\over 2}
+
{{-{{D}^2}}\over 6} + {{3\,{F^2}}\over 2}
,\nonumber\\ 
&&\beta^{4+i5}_{p\Lambda,ss}=
{1\over 4}
, 
\end{eqnarray}
\begin{eqnarray}
&&\beta^{4+i5}_{\Lambda\Xi^{-},uu}=
{1\over 4}
+
{{2\,{{D}^2} - 5\,D\,F + 3\,{F^2}}\over 3}
+
{{4\,{{C}^2}\,\left( -2\,D + 3\,F \right) }\over 
   {9\,\left( -D + 3\,F \right) }}
,\nonumber\\ 
&&\beta^{4+i5}_{\Lambda\Xi^{-},us}=
-{1\over 2}
+
{{-{{D}^2} - 10\,D\,F + 15\,{F^2}}\over 6}
+
{{2\,{{C}^2}\,\left( 5\,D - 3\,F \right) }\over 
   {9\,\left( -D + 3\,F \right) }}
,\nonumber\\ 
&&\beta^{4+i5}_{\Lambda\Xi^{-},ss}=
{1\over 4}
+
{{F\,\left( D + 3\,F \right) }\over 3}
+
{{2\,{{C}^2}\,\left( D + 3\,F \right) }\over 
   {9\,\left( D - 3\,F \right) }}
, 
\end{eqnarray}
\begin{eqnarray}
&&\beta^{4+i5}_{n\Sigma^{-},uu}=
{1\over 4}
+
F\,\left( -D + 3\,F \right) 
+
{{2\,{{C}^2}\,\left( D - 3\,F \right) }\over 
   {9\,\left( D - F \right) }}
,\nonumber\\ 
&&\beta^{4+i5}_{n\Sigma^{-},us}=
-{1\over 2}
+
{{{{D}^2} - 4\,D\,F + 3\,{F^2}}\over 2}
+
{{2\,{{C}^2}\,\left( D - 3\,F \right) }\over 
   {9\,\left( -D + F \right) }}
,\nonumber\\ 
&&\beta^{4+i5}_{n\Sigma^{-},ss}=
{1\over 4}
, 
\end{eqnarray}
\begin{eqnarray}
&&\beta^{4+i5}_{\Sigma^{0}\Xi^{-},uu}=
{1\over 4}
+
F\,\left( -D + F \right) 
+
{{2\,{{C}^2}\,\left( -9\,D - 9\,F + 5\,H \right) }\over 
   {81\,\left( D + F \right) }}
,\nonumber\\ 
&&\beta^{4+i5}_{\Sigma^{0}\Xi^{-},us}=
-{1\over 2}
+
{{{{D}^2} - 2\,D\,F + 5\,{F^2}}\over 2}
+
{{4\,{{C}^2}\,\left( 9\,D + 9\,F - 5\,H \right) }\over 
   {81\,\left( D + F \right) }}
,\nonumber\\ 
&&\beta^{4+i5}_{\Sigma^{0}\Xi^{-},ss}=
{1\over 4}
+
F\,\left( -D + F \right) 
+
{{2\,{{C}^2}\,\left( -9\,D - 9\,F + 5\,H \right) }\over 
   {81\,\left( D + F \right) }}
, 
\end{eqnarray}
\begin{eqnarray}
&&\beta^{4+i5}_{\Sigma^{+}\Xi^{0},ab}=\beta^{4+i5}_{\Sigma^{0}\Xi^{-},ab}
, 
\end{eqnarray}
\begin{eqnarray}
&&\beta^{8}_{pp,uu}=
{{{{\left( -D + 3\,F \right) }^2}}\over 2}
,\,\,\,
\beta^{8}_{pp,us}=\beta^{8}_{pp,ss}=
0
. 
\end{eqnarray}

\vspace{5mm}

\leftline{\bf\Large Acknowledgements}
We are thankful to T.-S. Park and P. Ko for useful discussions.
This work was supported by the Korean Science and Engineering
Foundation through the Center for Theoretical Physics.

\pagebreak

\end{document}